\documentclass[aps,prb,twocolumn,superscriptaddress]{revtex4}
\usepackage{amsmath}
\usepackage{amsfonts}
\usepackage{amssymb}
\usepackage{epsfig}
\usepackage{graphicx}

\begin{document}

\title{Non-monotonic temperature evolution of dynamic correlations 
in glass-forming liquids}

\author{Walter Kob}
\affiliation{Laboratoire Charles Coulomb,
UMR CNRS 5221, Universit{\'e} Montpellier 2,
34095 Montpellier, France}

\author{S\'andalo Rold\'an-Vargas} 
\affiliation{Laboratoire Charles Coulomb,
UMR CNRS 5221, Universit{\'e} Montpellier 2,
34095 Montpellier, France}
\affiliation{Departamento de Fisica
Aplicada, Grupo de Fisica de Fluidos y Biocoloides, Universidad de
Granada, 18071 Granada, Spain}

\author{Ludovic Berthier}
\affiliation{Laboratoire Charles Coulomb,
UMR CNRS 5221, Universit{\'e} Montpellier 2,
34095 Montpellier, France}

\date{\today}

\begin{abstract}

The viscosity of glass-forming liquids increases by many orders
of magnitude if their temperature is lowered by a mere factor
of 2-3~\cite{reviewnature,binder_book_10} .  Recent studies
suggest that this widespread phenomenon is accompanied by spatially
heterogeneous dynamics~\cite{ediger,reviewchi4}, and a growing dynamic
correlation length quantifying the extent of correlated particle
motion~\cite{glotzer_99,berthier_05,dalle_07}.  Here we use a novel
numerical method to detect and quantify spatial correlations which
reveal a surprising non-monotonic temperature evolution of spatial
dynamical correlations, accompanied by a second length scale that grows
monotonically and has a very different nature. Our results directly unveil
a dramatic qualitative change in atomic motions near the mode-coupling
crossover temperature~\cite{gotze_book}which involves no fitting or
indirect theoretical interpretation.  Our results impose severe new
constraints on the theoretical description of the glass transition,
and open several research perspectives, in particular for experiments,
to confirm and quantify our observations in real materials.

\end{abstract}

\maketitle

More than forty years ago Adam and Gibbs~\cite{adam_65} put forward
the seminal idea that the relaxation dynamics of highly viscous
liquids occurs through `cooperatively relaxing regions'. Their theory
suggested that particle motion occurs in a collective manner in  localized
domains~\cite{adam_65} whose typical size is related to the entropy of the
systems and increases with decreasing temperature. This result implied
that the relaxation mechanism is controlled by a unique length scale
of thermodynamic origin with a direct signature in the dynamics. This
approach regained momentum in the 1990's when novel experimental
techniques and large scale computer simulations established the presence
of dynamical heterogeneities, i.e. localized regions where dynamics
is significantly different from the average~\cite{ediger,reviewchi4},
although these observations can also be interpreted as a purely dynamical
phenomenon~\cite{chandler}.

A qualitatively similar, but much more detailed, theoretical description
is obtained within the framework of the random first order transition
(RFOT) theory~\cite{rfot}. Within this approach, there exists an ideal glass
transition that underlies glass formation, with an associated diverging
correlation length scale of entropic origin since it is related to the
existence of a large number of long-lived metastable states. At very low
temperatures, the glass-former is described as a `mosaic' of correlated
domains that rearrange in a thermally activated, collective manner,
such that again static and dynamic correlations coincide and grow
with the viscosity~\cite{rfot2}. However, models with an RFOT (mainly
mean-field like models) also display a `spinodal' singularity at a
higher temperature, $T_c$, at which metastability is lost and therefore
the mosaic picture is no longer useful~\cite{rfot,franz_07}. Thus
above $T_c$, a different approach must be used and it is found
that the dynamics within mean-field models~\cite{thirumalai} has
profound similarities with the one predicted from mode-coupling
theory~\cite{gotze_book}, where $T_c$ corresponds to a
dynamic critical point associated with a diverging dynamical correlation
length~\cite{BBepl}.  Thus, even at the theoretical level, the physics
around $T_c$ remains `mysterious'~\cite{BBreview}: How can a dynamical
correlation length diverge at two distinct temperatures?  The mystery
thickens in finite dimensions for which the mode-coupling singularity is
cut off, and its existence can be inferred only from fitting relaxation
data~\cite{gotze_book}, a procedure that is prone to criticisms, and
therefore the physical relevance of the `avoided' singularity at $T_c$
has remained a debated issue.

Progress is also slow because experiments on molecular systems
do not have enough resolution to follow atomic motions over long
times~\cite{ediger}, and numerical simulations often cannot access low
enough temperatures to make definite statements on the size and nature of
dynamic heterogeneities (in fact there are so far no numerical studies
on the dynamical heterogeneities below $T_c$).  At present, the most
direct measurements~\cite{reviewchi4} seem to indicate that the
dynamical correlation length increases from 1-2 particle diameters at
moderately supercooled temperatures to 5-10 diameters close to the glass
transition temperature, but the interpretation of the experimental data is
often rather difficult~\cite{dalle_07}. In this article we show that the
use of a new methodology to characterize bulk dynamic correlations reveals
details on their temperature evolution not observed in previous work.

In parallel to the quest for a dynamical length scale, evidence has
also been found for an increasing of static correlations. However,
this information is not captured by standard two-point
correlation functions. Recent work has for instance suggested
the growth of locally favored geometric structures in some model
systems~\cite{coslovich_07,tanaka}, but these methods are not
easily generalized to different glass-formers. One possibility
to tackle this problem are the point-to-set correlations which
are an elegant, general method to capture the multi-point static
correlations which might characterize the non-trivial structure of
viscous liquids~\cite{pointtoset}.  The conceptual idea is to `pin'
the position of a number of particles (the `set') in an equilibrated
configuration of the fluid, and to measure how the position of the
remaining particles is affected. It has recently been argued that
in the geometry in which particles outside a spherical cavity are
pinned, this point-to-set correlation should detect the typical domain
size of the RFOT mosaic state~\cite{BB2}.  Numerical simulations
confirmed qualitatively the growth of point-to-set correlations in
this particular geometry~\cite{biroli_08}.  However, the connection
to dynamic correlations and the precise temperature dependence of the
various length scales were not studied, and these results did not resolve
the `mystery'~\cite{BBreview} of the $T_c$ crossover.

Inspired by previous work on confined fluids~\cite{scheidler} 
(see also Ref.~\cite{watanabe11}), we
have generalized the idea of a point-to-set correlation to a novel
geometry. We pin particles in a semi-infinite space and detect the
resulting effect on the other half space.  The principal advantage is that
we can measure simultaneously the static and dynamic profiles induced by
the frozen wall.  Additionally we are able to perform  simulations from the
high temperature liquid down to and below $T_c$ with a realistic molecular
dynamics, thus allowing us to resolve at once multi-point static and
dynamic correlations in a very broad temperature regime encompassing the
(hypothetical) mode-coupling crossover.

We study a binary mixture of quasi-hard spheres~\cite{tom}, as described
in the {\it Methods}. The fluid is equilibrated by means of standard
molecular dynamics using periodic boundary conditions in all three
directions. To simulate particles pinned within a semi-infinite space,
$z<0$, it is enough to freeze at an arbitrary time $t=0$ the position
of all particles within a slice of thickness $d_{\rm wall}=1.4 \sigma$
which is perpendicular to the $z-$axis: They form our `set'. 
Because we use periodic boundary conditions, we
work with a very large system size in the $z$-direction, ensuring
that bulk behavior is recovered at the center of the simulation
box, i.e. the replicated walls do not interfere with each other.

\begin{figure}
\includegraphics[scale=0.32,angle=0,clip]{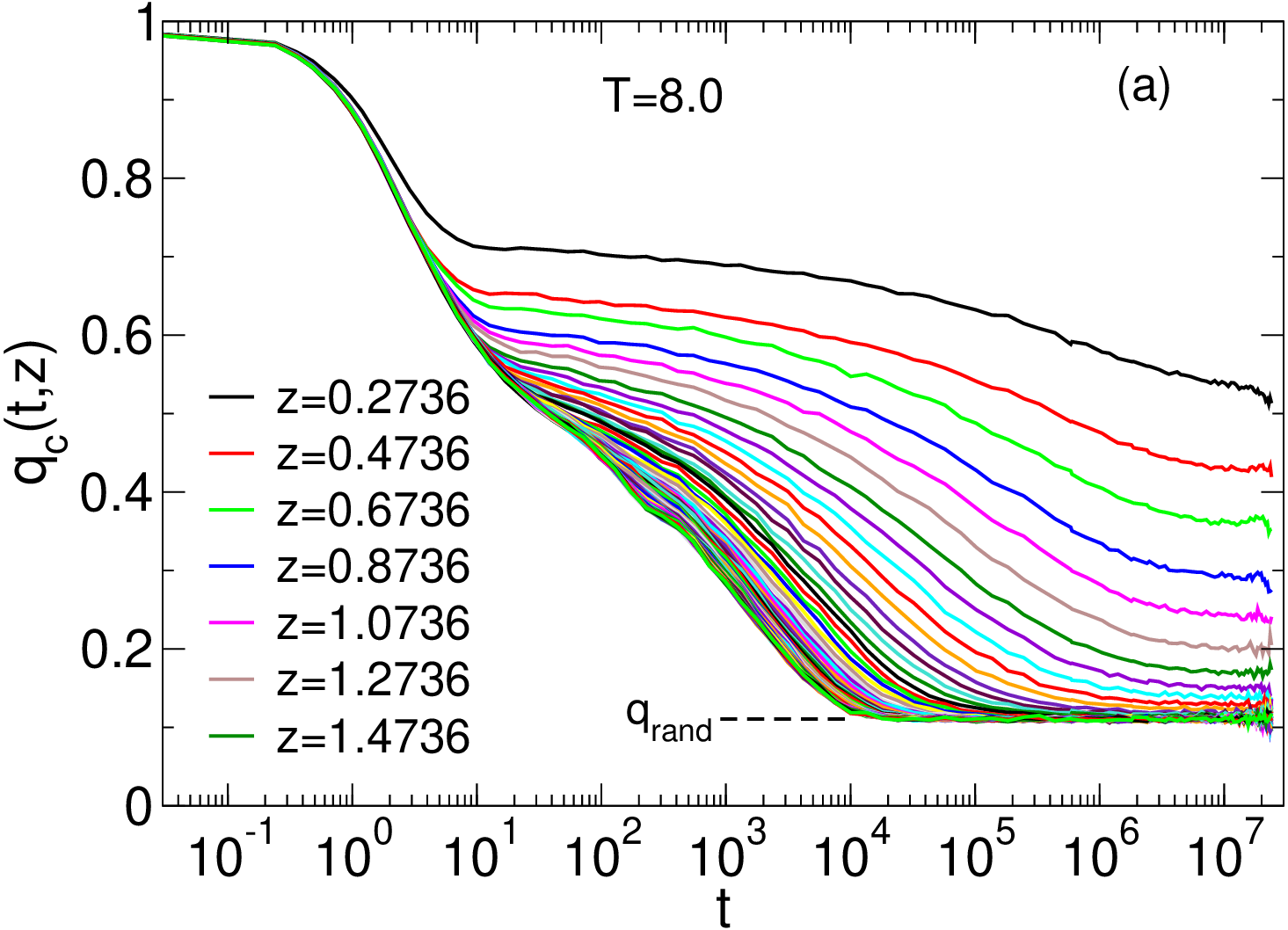}
\includegraphics[scale=0.32,angle=0,clip]{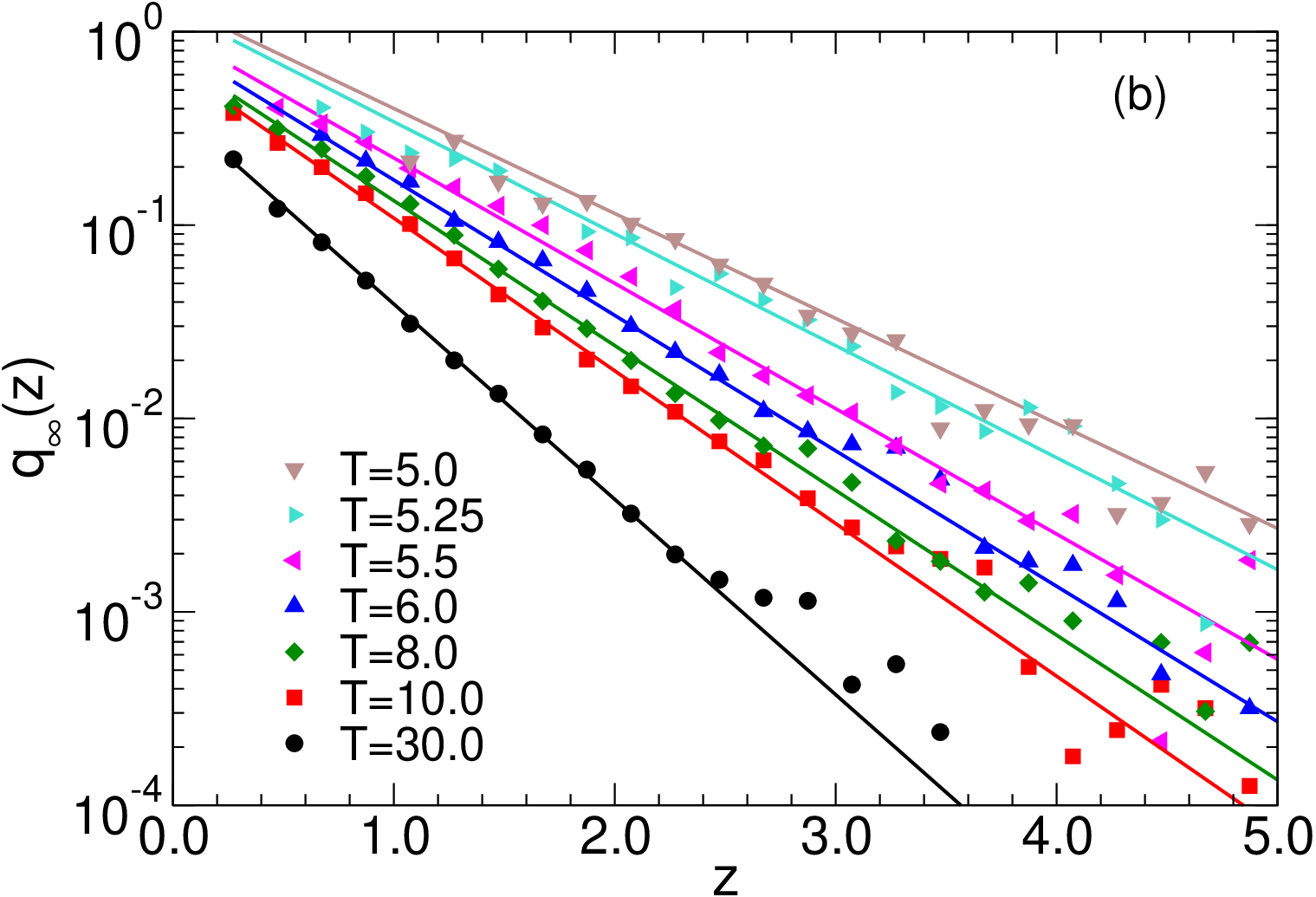}
\caption{(a) Time dependence of the overlap $q_c(t,z)$, Eq.~(\ref{eq1}),
for $T=8.0$ and various values of $z$, increasing from $z=0.2736$ (near the
wall, rightmost curve) by increments of 0.2. The leftmost curve presents
the bulk behavior.  (b) $z-$dependence of the nontrivial part of the
static overlap $q_\infty(z)$ for different temperatures (symbols). The
solid lines are fits with the exponential given in Eq.~(\ref{eq3}).}
\label{fig2}
\end{figure}

To properly measure a point-to-set correlation it is crucial that the
frozen walls have the same structure as the equilibrated liquid at
temperature $T$, such that the average static properties of the confined
liquid are unperturbed~\cite{pointtoset,scheidler} (see also {\it Supplementary
Material}).  We have measured how far (in $z$) the wall influences 
the static local density field and its dynamics, giving us independent
access to static and dynamic correlation length scales. Note that within
RFOT theory, the spatial extent of the static profile near a wall is
not directly controlled by the mosaic length scale~\cite{franz_07}, and a
comparison with results obtained with a spherical cavity~\cite{biroli_08}
is not straightforward.

A convenient observable to characterize the influence of the wall is
the overlap profile $q_{c}(t,z)$, defined as follows~\cite{biroli_08}.
We discretize space into small cubic boxes of linear size $\delta \approx 0.55
\sigma$, and define $n_i(t)=1$ if box $i$ is occupied by at least one
particle at time $t$, and $n_i(t)=0$ if not.  The overlap profile in the $z$
direction with respect to the template configuration at time $t=0$ is

\begin{equation}
q_{c}(t,z)=\left[\frac{\sum_{i(z)} \langle n_i(t) n_i(0) 
\rangle}{\sum_{i(z)} \langle n_i(0)\rangle}\right]_{\rm wall} \quad ,
\label{eq1}
\end{equation}

\noindent
where the sums run over all boxes at distance $z$ from the wall, $\langle
\cdots \rangle$ is the usual thermal average, and $[ \cdots ]_{\rm
wall}$ is an additional average over independent wall realizations.
Thus $q_{c}(t,z)$ quantifies the similarity of particle configurations
separated by a time $t$ at distance $z$, and, by construction,
$q_{c}(t=0,z) = 1$, for all $z$.  We have also studied $q_s(t,z)$, the
single particle version of Eq.~(\ref{eq1}), obtained by requesting that
the box $i$ is occupied at times 0 and $t$ by the {\it same} particle.
We find no relevant difference between these two correlation functions
as far as time dependence is concerned.

In Fig.~\ref{fig2}a we show the time dependence of $q_c(t,z)$ at
$T=8.0$ for different values of $z$. For large $z$ (leftmost curves)
the correlators become independent of $z$ and present the bulk behavior.
The presence of a shoulder at intermediate times reflects the usual cage
motion of particles observed in glassy systems~\cite{binder_book_10}. In
the long-time limit, $t \to \infty$, the correlator decays to $q_{\rm
rand} = 0.110595$, the probability that a box is occupied, a quantity
we measured with high precision from bulk simulations.  With decreasing
$z$, the height of the plateau at intermediate times increases and the
timescale to reach it decreases. The final decay is much slower near
the wall than in the bulk, showing that the $\alpha$-relaxation is
strongly affected by the frozen wall. Furthermore the long-time limit
of the overlap increases from its trivial value, $q_c(t\to\infty,z)
> q_{\rm rand}$, showing that sufficiently close to the wall, the
density cannot freely fluctuate. (See {\it Supplementary Material} for
an illustration.) Thus the set of frozen particles with $z<0$ influences
the position of the liquid particles at $z>0$ over a non-trivial static
length scale.  Finally, Fig.~\ref{fig2}a shows that there exists a range
of $z$ values for which the long-time limit of the overlap is the trivial
bulk value, while the relaxation timescale is slower than the bulk. This
directly shows, with no further analysis, that dynamic correlations have
a larger range than static ones, as we confirm below.

To quantify these observations, we fit the final decay of
$q_c(t,z)$ to the stretched exponential form
\begin{equation}
q_c(t,z) - q_{\rm rand} = A \exp[-(t/\tau)^{\beta}]+q_\infty, 
\label{eq2}
\end{equation}
%
where $A$, $\tau$, $\beta$, and $q_\infty$ are fitted for each $z$. The
profile of the static overlap, $q_\infty(z)$ measures how far from the
wall density fluctuations are correlated, while $\tau(z)$ measures how
far dynamics is affected.  We find that Eq.~(\ref{eq2}) also describes
well the single particle overlap, $q_s(t,z)$, with the obvious difference
that $q_s(t\to\infty,z)=0$, because eventually all particles leave the
box that they occupy at $t=0$.  Thus, we obtain a second, `self' dynamic
profile from the study of $q_s(t,z)$.

In Fig.~\ref{fig2}b, we display the temperature evolution of the static
overlap profiles $q_\infty(z)$.  The semi-log plot suggests to describe
these data using an exponential decay~\cite{parisi_09}
\begin{equation}
q_\infty(z)= B \exp(-z/\xi^{\rm stat}),
\label{eq3}
\end{equation}
%
which allows us to define a static point-to-set correlation length scale
$\xi^{\rm stat}(T)$. From these data it is clear that $\xi^{\rm stat}$
grows when temperature decreases, a result in good qualitative agreement
with previous work~\cite{biroli_08} using a very different
geometry.
Notice that $B(T)$ also changes rapidly with $T$, which suggests to define
$\xi^{\rm stat-int} \equiv B(T) \xi^{\rm stat}$ as a convenient estimate
of the integrated profile, $\xi^{\rm stat-int} \approx \int_0^\infty
q_\infty(z) dz$.

\begin{figure}
\includegraphics[scale=0.32,clip]{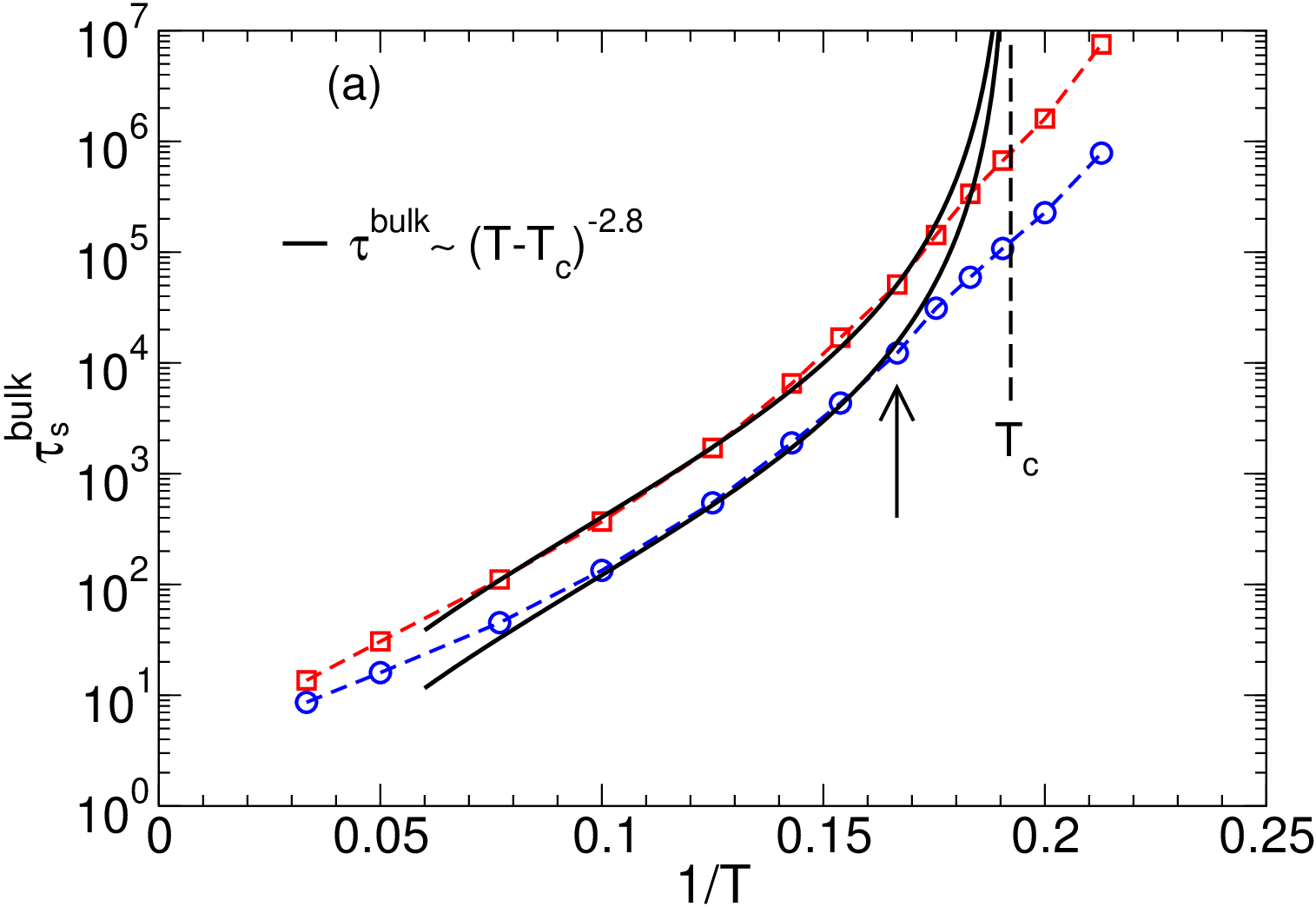}
\includegraphics[scale=0.32,clip]{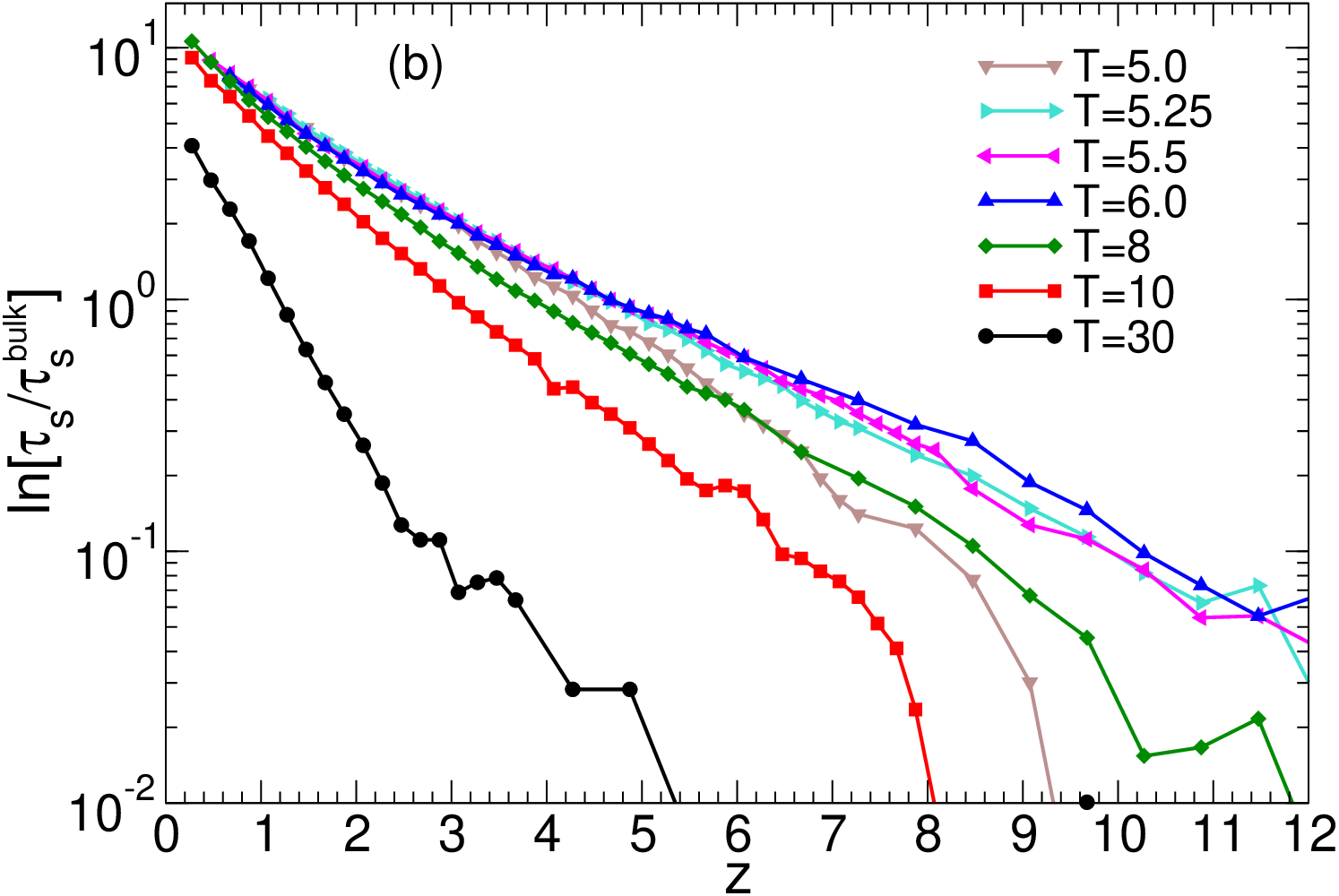}
\caption{(a) Temperature evolution of bulk relaxation time for large
(top) and small (bottom) particles. Solid lines represent a power-law
fit with $T_c=5.2$ (dashed line), the arrow indicates $T=6.0$ where
deviations from the fit appear.  (b) $z-$dependence of the relaxation
time $\tau_s(z,T)$ at various temperatures. Note the non-monotonic
$T-$dependence at intermediate and large $z$.}
\label{fig3}
\end{figure}

We now analyze the dynamic profiles. To take into account the fact
that the amplitude and stretching of the time dependent correlations
evolve with $z$, see Fig.~\ref{fig2}a, we have calculated the area
under the correlators $q_c(t,z)$ and $q_s(t,z)$, taking into account
only the secondary, slowly relaxing part.  We denote the resulting
times by $\tau_c(z)$ and $\tau_s(z)$, respectively. Previous
studies~\cite{scheidler} have suggested that for large $z$ the
$z-$dependence of $\tau_s(z)$ can be described well by an exponential
functional form,
\begin{equation}
\log(\tau_s)= \log(\tau_s^{\rm bulk}) + B_s \exp(-z/\xi_s^{\rm dyn}),
\label{eq4}
\end{equation}
%
where $B_s(z)$ and, more importantly, the dynamic length scale,
$\xi_s^{\rm dyn}(T)$, are adjusted for each $z$.  The bulk relaxation
time $\tau_s^{\rm bulk}(T)$ is measured independently with a very
good precision, see Fig.~\ref{fig3}a. Using a power law fit inspired
by mode-coupling theory~\cite{gotze_book}, $\tau^{\rm bulk}(T) \sim
(T-T_c)^{-\gamma}$, we obtain $T_c \approx 5.2$, but deviations from
the algebraic fit appear above $T_c$ near $T=6.0$, see Fig.~\ref{fig3}a
(see {\it Supplementary Material} for a discussion of this fit.)

In Fig.~\ref{fig3}b we show that the data at large $z$ can indeed be
fitted well by the Ansatz given by Eq.~(\ref{eq4}). We have found a similar
behavior for $\tau_c(z)$, providing us with a second dynamic correlation
length scale, $\xi_c^{\rm dyn}$.  For high temperatures Eq.~(\ref{eq4})
gives a good description of the data over the entire range of distances
$z$.  For intermediate and low temperatures we see the development of
a curvature in a semi-log plot, indicating at small $z$ deviations
from the simple exponential dependence. This might suggest, although very
indirectly, the appearance of more than one relaxation process for
the relaxation dynamics. 

A remarkable behavior occurs at intermediate and large distances,
which has, to our knowledge, remained undetected. The dynamic profiles
exhibit a striking non-monotonic evolution with temperature.  A close
inspection of Fig.~\ref{fig3}b shows that the dynamic profiles extend to
increasingly larger distances when temperature decreases from $T=30.0$
down to $T=6.0$, but become shorter-ranged when $T$ is decreased further,
down to $T=5.5$, $T=5.25$ and then $T=5.0$. The maximum occurs near $T=6.0$,
which is also the temperature at which deviations from mode-coupling fits
appear, see Fig.~\ref{fig3}a.

We have carefully checked that this behavior is not a result of our
numerical analysis. A direct visual inspection of the time correlations
functions $q_s(z,t)$ reveals that the spread of the curves in the slow
decay has a maximum at $T=6.0$, so that the non-monotonic temperature
behavior in Fig.~\ref{fig3}a is not an artifact of our fits, but is a
genuine effect (see {\it Supplementary Material}, Fig. SM-2). In addition,
we found very similar results for the collective relaxation time,
$\tau_c(z,T)$, which further shows that this non-monotonic behavior does
not sensitively depend on the considered observable.  Thus, these results
give us direct evidence that the relaxation processes responsible for
spatial dynamic correlations have a non-monotonic temperature behavior.
To our knowledge, all previous numerical and experimental studies of
spatially heterogeneous dynamics have reported spatial correlations
which grow as the temperature is decreased and dynamics slows
down~\cite{ediger,reviewchi4,glotzer_99,berthier_05,dalle_07,tanaka}.

\begin{figure}
\includegraphics[width=8.5cm,angle=0,clip]{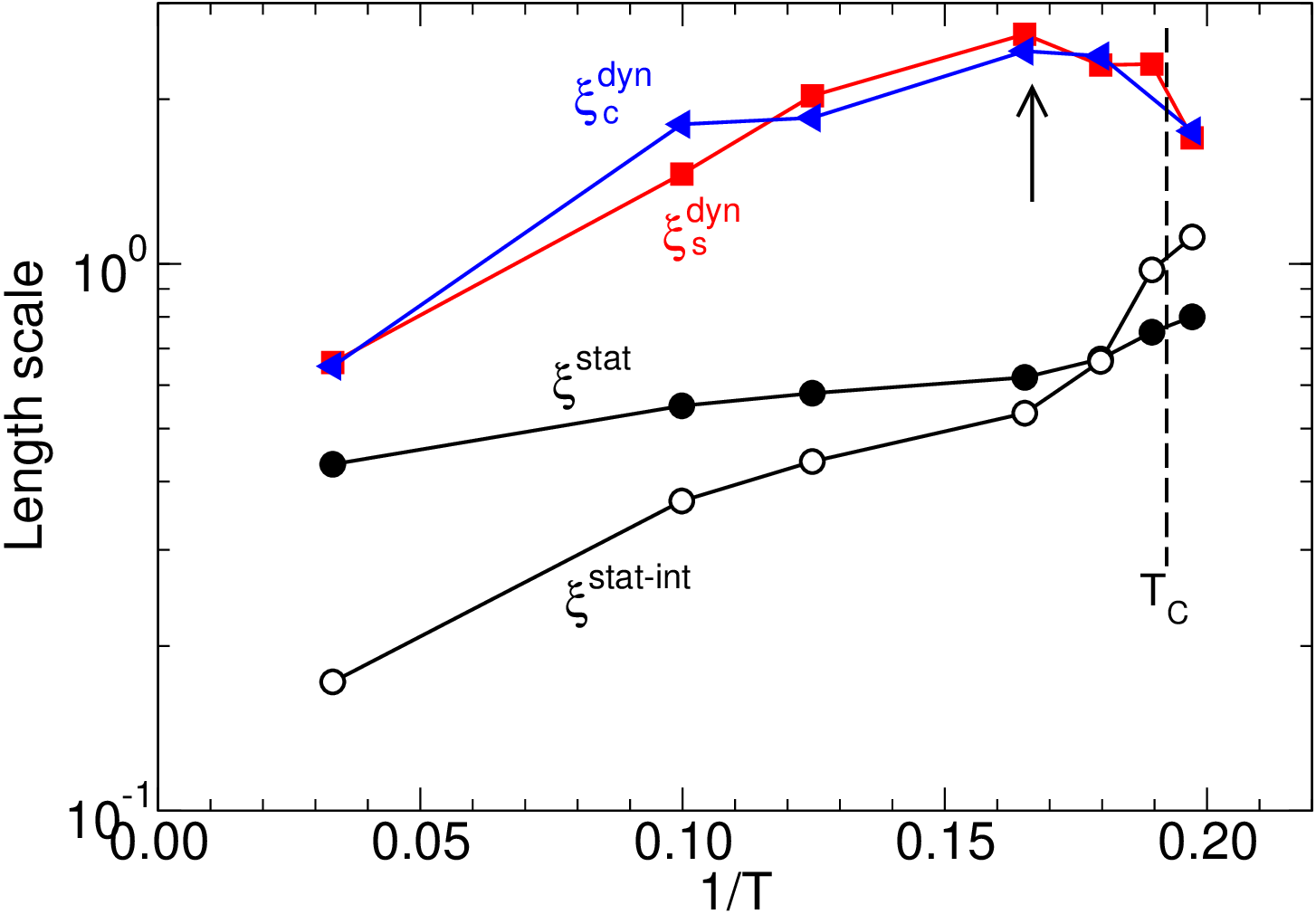}
\caption{Temperature dependence of static and dynamic length
scales identified in this work. The arrow at $T=6.0$ and 
the dashed line at $T_c=5.2$ are as in Fig.~\ref{fig3}.
Dynamic length scales display a non-monotonic behavior with a maximum 
at $T=6.0$, while static length scales increase modestly above $T_c$. }
\label{fig5}
\end{figure}

In Fig.~\ref{fig5} we summarize the temperature dependence of
the static and dynamic length scales identified above.  The static
length scales show a modest but steady and monotonic growth with
decreasing temperature, which seems to become more pronounced below
$T \approx 6.0$.  These are natural findings from the RFOT theory
perspective~\cite{rfot,franz_07,BBreview}, in the sense that static
correlations should only become prominent below $T_c$.  We also include
in Fig.~\ref{fig5} the dynamic length scales, which have a striking
local maximum near $T=6.0$.  A comparison with Fig.~\ref{fig3}a
strongly suggests to interpret this maximum in the context of RFOT
theory in which a dynamic critical point at $T_c$ exists in the
mean-field limit (but which is avoided in finite dimensions) whereas
activated dynamics governed by growing static correlations appears at
low temperatures. Using RFOT theory Stevenson {\it et al.} have suggested
that around $T_c$ the cooperative domains should indeed change shape, in
that they have an open structure above $T_c$ and a more compact structure
below $T_c$~\cite{stevenson_06}.  Our findings may be viewed as a striking
confirmation of this scenario. Although a gradual change from `flow-like'
to `hopping' motion was often invoked in the past~\cite{gotze_book},
mainly to rationalize successes and failures of the mode-coupling theory,
our results provide a very direct, microscopic evidence of a change of
relaxation mechanism which involves no theoretical fitting or indirect
interpretation.

Although the presented results have been obtained for a system in
the presence of pinned disorder, we emphasize that they reflect the
behavior of the liquid in the {\it bulk}, since the nature of the boundary
condition does not affect averaged static properties.  No non-monotonic
dynamic length scale has been detected so far in bulk systems (leaving out
systems that show also anomalies in their thermodynamics), but we think
that this is only related to the fact that previous measurements (such
as four point correlation function $\chi_4$~\cite{ediger,reviewchi4})
only provide a coarse representation of spatial correlations, since
they mainly probe the total number of `fast relaxing particles' and not
the details of the shape of the relaxing regions~\cite{reviewchi4}.
Therefore the present results should not be taken as a contradiction
to previous experiments and simulations, but as a new insight into
the relaxation mechanism. We hope that in the future new experimental
techniques will allow to detect the non-monotonic $T-$dependence of
$\xi^{\rm dyn}$ also in real molecular systems.

If we naively extrapolate our results to lower temperatures, we find a
temperature below which static correlations become larger-ranged than
dynamic ones, a situation which is physically not very meaningful. Thus we are
led to speculate that at much lower temperatures, dynamic length scales
should exhibit an upturn, and perhaps become slaved to the static ones,
as in the Adam-Gibbs picture~\cite{adam_65} and the scaling regime
of RFOT theory~\cite{rfot}.  However, studying numerically this final
regime is at present too difficult.  We suggest that experimental work
is needed to resolve these issues further. Our study also suggests
that investigations of confined systems should be revisited in both
simulations and experiments, and the mode-coupling crossover studied more
extensively in glassformers with different fragility. However, since the
model investigated here has no unusual features regarding the relaxation
dynamics~\cite{tom}, we expect our results to apply also to other simple
models such as hard spheres, Lennard-Jones-like systems, or soft spheres.

{\bf Methods--} We study an equimolar binary mixture of harmonic
spheres~\cite{tom} with diameter ratio 1.4, and interactions between
particle $i$ and $j$ given by

\begin{equation}
V_{ij}(|\vec{r}_i - \vec{r}_j|)= 
\frac{\varepsilon}{2} (1-|\vec{r}_i - \vec{r}_j|^2/\sigma_{ij}^2)^2 \quad 
\mbox{if } \quad |\vec{r}_i - \vec{r}_j| < \sigma_{ij} \quad ,
\label{eq6}
\end{equation}

\noindent
where $\sigma_{11} \equiv \sigma$ is the unit of length,
$\sigma_{12}=1.2$, and $\sigma_{22}$=1.4.  The total number of particles
is 4320 and all of them have the same mass $m$. Time is expressed in units
of $\sqrt{m \sigma^2/\varepsilon}$ and temperature in units of $10^{-4}
\varepsilon$, setting the Boltzmann constant $k_B=1.0$. We have used a
rectangular box of size $L_x=L_y=13.68$ and $L_z=34.2$, yielding a number
density $\rho = 0.6749715$. This system size is sufficiently large to avoid finite 
size effects.
The equations of motion have been integrated
with the velocity form of the Verlet algorithm. The longest runs extended
over 830 million time steps, which took about 6 weeks of CPU time on a
high end processor.  In order to improve the statistic of the results we
have averaged over 10-30 independent walls.  The total amount of computer
time to obtain the described results was therefore around 7 years. In
practice, to prevent particles to penetrate in the frozen half-space,
we introduced at the two surfaces of the frozen slice an infinitely
hard wall, and have checked that this frozen geometry has negligible
influence on the structure of the fluid. (See also {\it Supplementary
Material}.) Note that we are investigating here a dynamical equilibrium,
i.e. all fluid particles can leave their initial positions and explore
the whole confined space, thus assuring thermodynamic equilibrium conditions.

{\bf Acknowledgments}
We thank G. Biroli and A. Cavagna for fruitful exchanges about this work,
and the R\'egion Languedoc-Roussillon (L.B.), ANR DYNHET (L.B. and W.K.),
and MICINN (Project: MAT2009-13155-C04-02) and Junta de Andaluc\'{i}a
(Project: P07-FQM-02496) (S.R.V.) for financial support. W.K. is member
of the Institut universitaire de France.

\newpage

\begin{center}
{\Large \bf Supplementary Material}
\end{center}

\noindent

{\bf Influence of the amorphous wall on the local structure of the system}

In order to understand better how the presence of the amorphous wall
influences the local structure of the confined liquid we show in Fig.-SM-1
a slice of the system orthogonal to the confining walls. The two panels
correspond to the two temperature $T=30$ and $T=4.5$.

This figure was obtained by making a run that is about 100 times longer
than the $\alpha-$relaxation time in the bulk (=$\tau_s^{\rm bulk}(T)$)
and by superimposing snapshots of the location of the particles every
$\tau_s^{\rm bulk}$. In order to avoid overcrowding, only the particles
in a slice of thickness $\Delta x=1.0$ are shown. From this figure one
can see that in the center of the box the snapshots start to fill up the
space, i.e. the density of the particles becomes uniform, as it should
be in a bulk liquid. In contrast to this, the presence of the walls
influences the {\it local} density. However, as one can demonstrate
analytically, see Ref. [27] {\it the structure averaged parallel to
the wall is independent of $z$}, i.e. is the one of the bulk if one
has i) either an infinitely extended wall, or ii) makes an average over
sufficiently many independent realizations of the wall.

Comparing panels a) and b) we see that the range over which one finds an
influence of the amorphous walls on the local structure of the confined
liquid does indeed grow with decreasing temperature, and in the manuscript
we demonstrate that this influence decreases exponentially with $z$ with a
temperature dependent prefactor (figure 1b). To quantify the distance
over which the wall (locally) modifies the density profile, it is
reasonable to use $\xi^{\rm stat-int}$, instead of $\xi^{\rm stat}$,
since the former length takes into account also the prefactor of the
mentioned exponential.\\[10mm]

\begin{figure}
\includegraphics[scale=0.34,angle=0,clip]{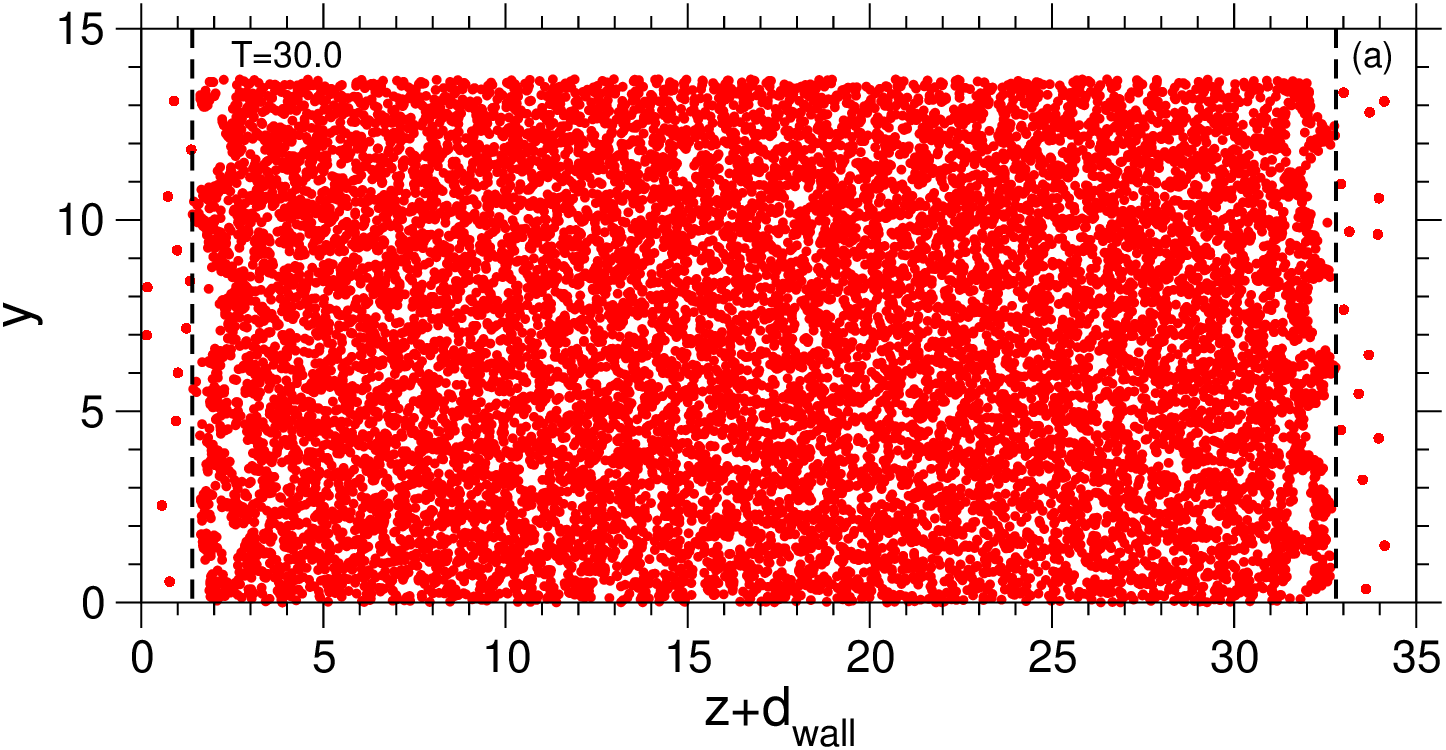}
\includegraphics[scale=0.32,angle=0,clip]{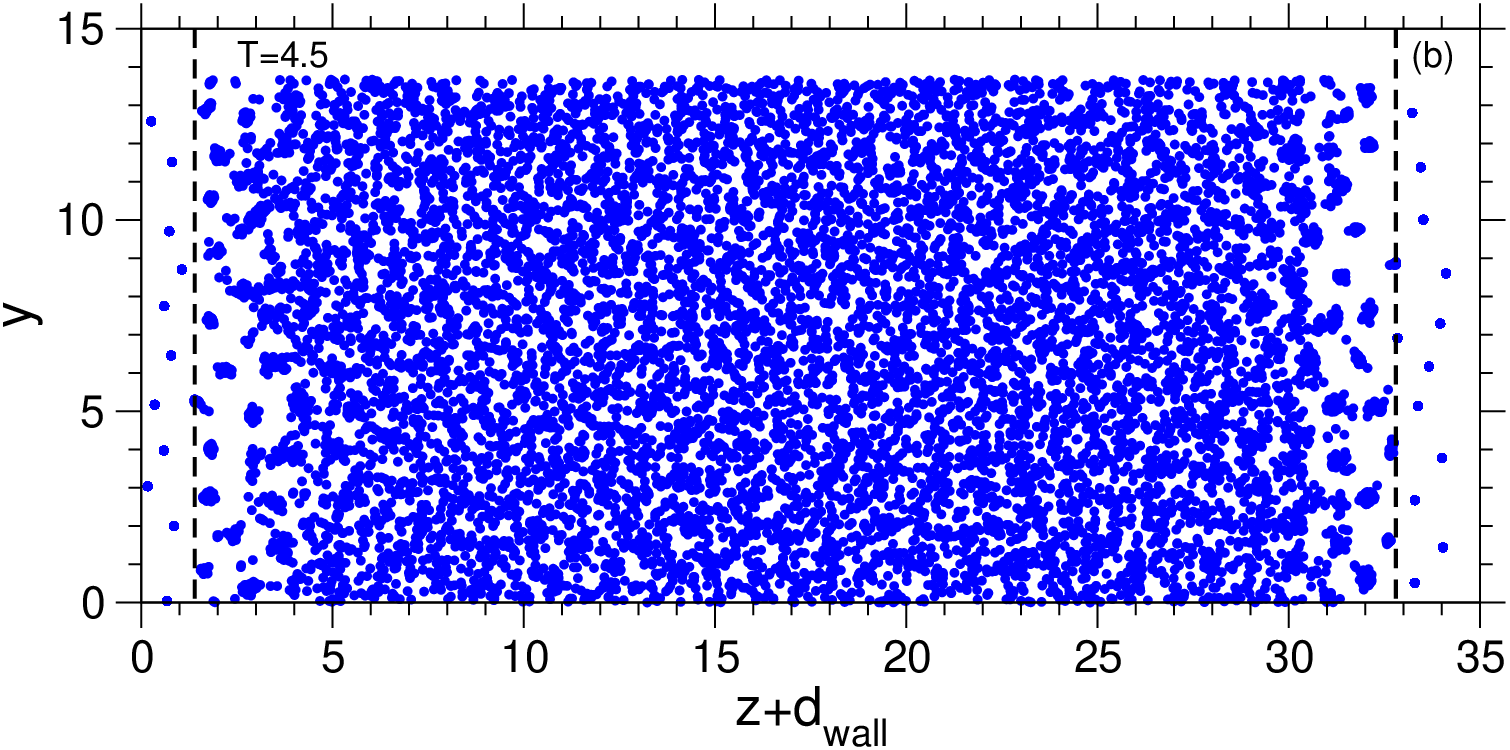}
\caption{
Figure SM-1: Superposition of snapshots of the system at $T=30.0$ and
$T=4.5$ (panel (a) and (b), respectively). The vertical dashed lines
show the location of the hard walls.  Only particles in the slab $0\leq
x_i \leq 1.0$ are shown. Individual snapshots are separated by about
$\tau_s^{\rm bulk}(T)$ and the total time is about 100 $\tau_s^{\rm
bulk}(T)$.}
\end{figure}

{\bf Time dependence of the overlap at fixed distance from the wall}

In order to show that the non-monotonic temperature dependence of the
dynamic length scale $\xi_s^{\rm dyn}$ is  not just an effect of the
way we have extracted this length scale from the data, and that the
decrease is present at {\it several} temperatures, we show in Fig.~SM-2
the time dependence of the self-overlap $q_s(z,t)$ at fixed value of $z$
and several temperatures. One recognizes that at high temperatures, see
curves for $T=10$, the curves for the different values of $z$ almost
coincide, which implies that the length scale over which the dynamics
is influenced by the amorphous walls is very small. With decreasing
temperature, curves for $T=6.0$, one sees a significant spread of
the curves in the $\alpha-$regime, indicating that $\xi_s^{\rm dyn}$
has increased. It the temperature is lowered even further ($T=5.25$
and $T=5.0$) this spread is less pronounced than at $T=6.0$, i.e. the
$\xi_s^{\rm dyn}$ has decreased again. From this figure we thus can
conclude that the non-monotonic $T-$dependence of  $\xi_s^{\rm dyn}$ is
not an artifact of the way we have analyzed the data and that it is seen
directly from the time correlation functions at {\it all} temperatures
below $T=6.0$.\\[10mm]

\begin{figure}
\includegraphics[scale=0.32,angle=0,clip]{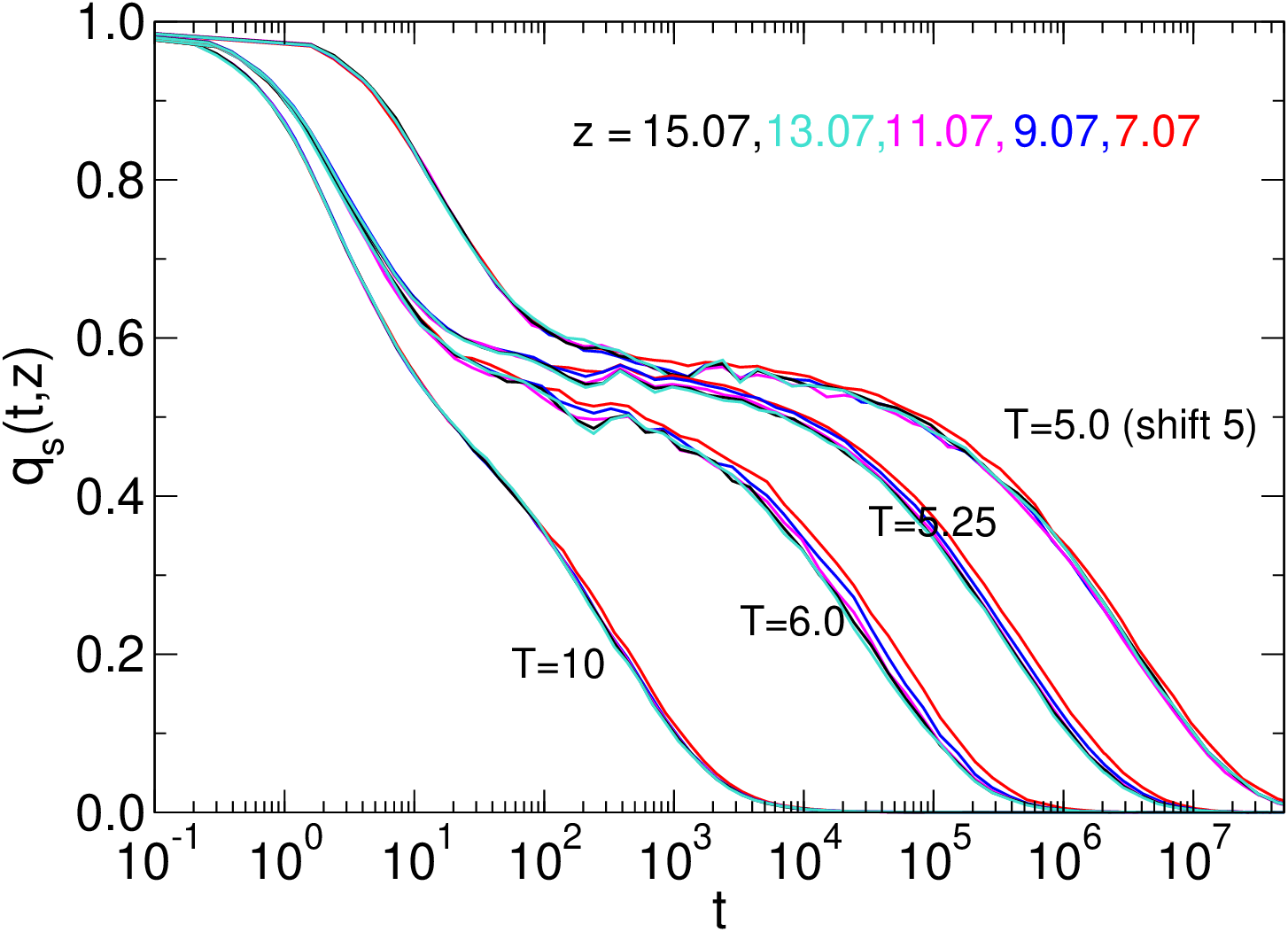}
\caption{Figure SM-2: Time dependence of the self-overlap $q_s(z,t)$ for selected
values of $z$. The temperatures are $T=10.0$, 6.0, 5.25, and 5.0. The
curves for $T=5.0$ have been shifted to the right by a factor of 5.0
in order to avoid overcrowding. Note that in the $\alpha-$relaxation
regime the spread of the curves is larger for $T=6.0$ than for the other
temperatures, thus showing the non-monotonic temperature dependence of
the length scale $\xi_s^{\rm dyn}(T)$. }
\end{figure}

\begin{figure}[t]
\includegraphics[scale=0.32,angle=0,clip]{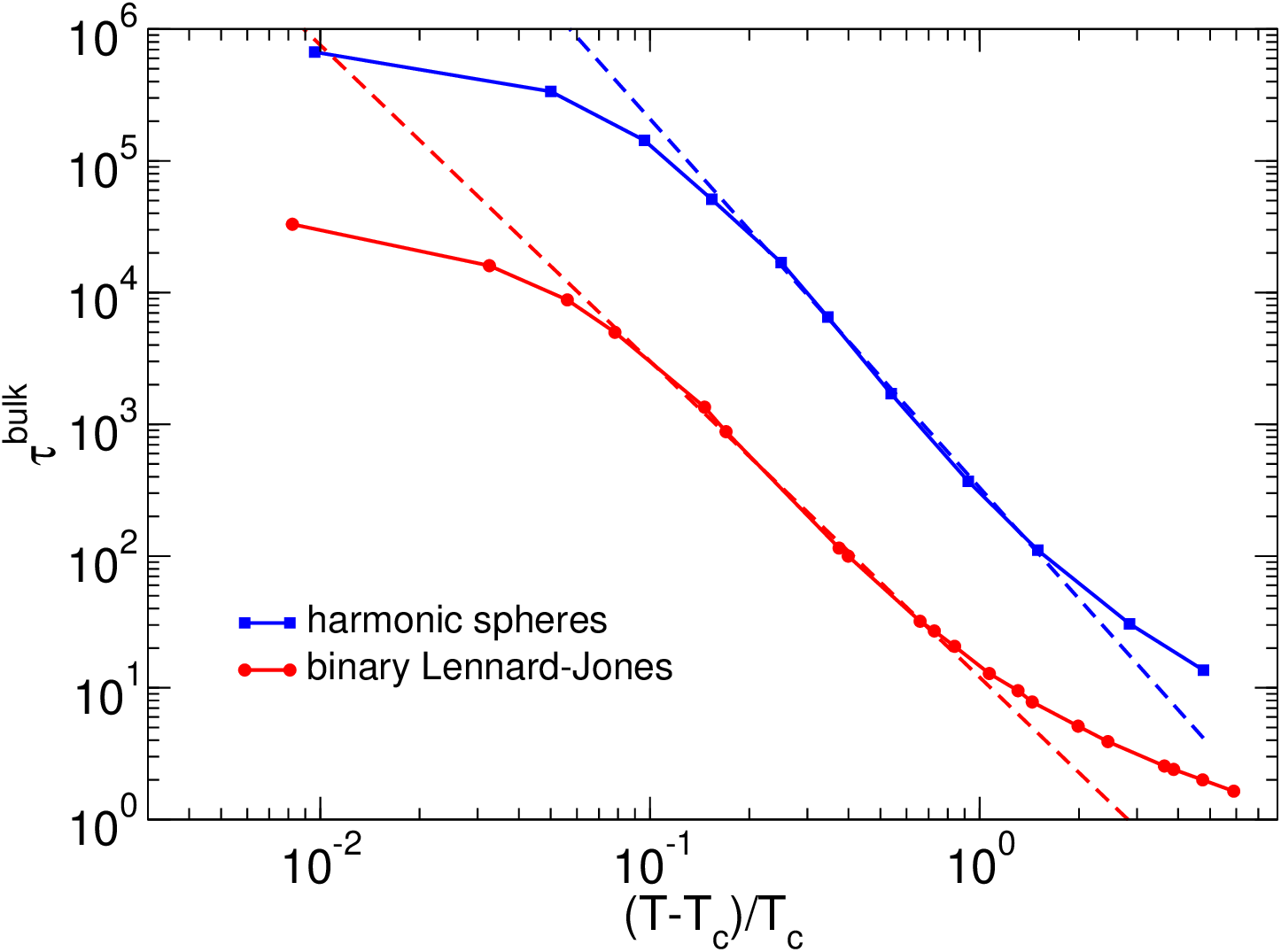}
\caption{Figure SM-3: Relaxation time for the bulk as a function of $T-T_c$ where
$T_c$ has been determined by trying to rectify the data. The squares
are for the large particles of the harmonic spheres and the circles for
the majority particles of the binary Lennard-Jones mixture from Ref.~[29].}
\end{figure}

{\bf Determination of the critical temperature of mode-coupling 
theory}

In Fig. 2a of the paper we show an Arrhenius plot of the relaxation
times for the system of harmonic spheres. Also included in this graph are
fits with the predicted power-laws of mode-coupling theory [28]. Following
standard practice in this field, these fits have been obtained in the
following way: One assumes a value of $T_c$ and makes a log-log plot of
$\tau_s^{\rm bulk}$ vs. $T-T_c$, using of course only data point with $T>
T_c$. Mode-coupling theory predicts that

\begin{equation}
\tau_s^{\rm bulk} \propto (T-T_c)^{-\gamma} \quad ,
\end{equation}

\noindent
where the exponent $\gamma$ is larger than 1.5, and has for
hard-sphere-like systems usually a value between 2.0 and 3.0~[8]. By
choosing the value of $T_c$ in such a way as to rectify the $\tau_s^{\rm
bulk}(T)$ data in a large range, one can determine $T_c$. In Fig.~SM-3
we show the result of such an adjustment for the large particles and we
see that the power-law describes the data over about three decades. We
find that for our system the exponent $\gamma$ is 2.8, i.e. a value
that is very similar to the ones that are found for other glass-forming
systems. For the sake of comparison we have included in the graph also the
data for the binary Lennard-Jones mixture of Kob and Andersen, a system
which has been found to be described very well by mode-coupling theory
[28,29]. The comparison between the data for the Lennard-Jones system with
the one from the harmonic spheres shows that the power-law of mode-coupling
theory describes the relaxation times well over the same number of decades
in $\tau_s^{\rm bulk}$. Therefore we can conclude that also for the system
of harmonic spheres the value of $T_c$ can be estimated with good accuracy.

We also emphasize that the results we presented here that concern the
self-quantity ($q_s(z,t)$ and $\tau_s^{\rm bulk}$) are confirmed by the
ones for the collective quantities ($q_c(z,t)$ and $\tau_c^{\rm bulk}$),
i.e. there is no relevant difference between these two observables as far as 
relaxation times are concerned.

\vspace*{20mm}
{\bf References}

[27] P. Scheidler, W. Kob, and K. Binder, {\it The relaxation dynamics
of a supercooled liquid confined by rough walls}, 
J. Phys. Chem. {\bf 108}, 6673 (2004).

[28] W.~Kob and H.~C. Andersen, 
{\em Testing mode-coupling theory for a supercooled binary
Lennard-Jones mixture: The van Hove correlation function}
Phys. Rev. E {\bf 51}, 4626 (1995).

[29] M. Nauroth and W. Kob, 
{\em A quantitative test of the mode-coupling theory of the ideal
glass transition for a binary Lennard-Jones system}
Phys. Rev. E {\bf 55}, 657 (1997).

\end{document}